\documentclass[twocolumn,secnumarabic,amssymb, nobibnotes, aps, prd]{revtex4-1}

\usepackage{amssymb,amsmath,amsfonts}
\usepackage{bbm,wasysym}
\usepackage{graphicx}
\usepackage{verbatim}
\newcommand{\id}{{\mathbbm{1}}}

\newcommand{\bra}[1]{{\langle #1 |}}
\newcommand{\ket}[1]{{| #1 \rangle}}

\newcommand{\tr}{tr}

\begin{document}

\title{Optimal super dense coding over memory channels}%

\author{Z. Shadman }%
\email[]{shadman@thphy.uni-duesseldorf.de}
\author{ H. Kampermann }{}
\author{D. Bru\ss}
\affiliation{ Institute f\"ur Theoretische Physik III,
              Heinrich-Heine-Universit\"at
              D\"usseldorf, D-40225 D\"usseldorf, Germany}
\author{C. Macchiavello}
\affiliation{ Dipartimento di Fisica  ``A. Volta" and INFM-Unit$\acute{a}$ di Pavia, Via Bassi 6, 27100, Pavia, Italy}

\date{\today}%

\begin{abstract}

We study the super dense coding capacity in the presence of 
  quantum channels with  correlated noise. We investigate  both the cases of
 unitary and non-unitary encoding.
Pauli
channels for arbitrary dimensions are treated explicitly.
The super dense coding capacity for some special channels and resource states 
is derived for unitary encoding. We also provide an example of a memory channel where non-unitary encoding leads to an improvement in the super dense coding capacity.

\begin{description}{}
\item[PACS numbers] 03.67.-a, 03.67.Hk, 03.65.Ud
\end{description}
\end{abstract}
\maketitle

\section{Introduction}

Super dense coding is one of the notable areas in which quantum
entanglement plays a crucial role. By this protocol, due to the
nonlocal properties of quantum entanglement, it is possible to
communicate two bits of classical information by sending one qubit
only \cite{Bennett}. The first attention, after proposing the super dense
 coding protocol, was given to various scenarios over noiseless channels
  and unitary encoding \cite{hiroshima, ourPRL, Dagmar}. In this case 
one starts from a $d^2$-dimensional bipartite shared state $\rho$ 
between the sender Alice and the receiver Bob. 
Alice performs with probability $p_i$ a local
   unitary operation $W_i$  on her subsystem to encode  classical
information through the state $\rho_i= (W_i\otimes\id)\rho({W_i}^{\dagger}\otimes\id).$ Subsequently, she sends her subsystem to Bob
(ideally via a noiseless channel). The
ensemble that Bob receives is $\{\rho_i,p_i\}$. The maximal amount
of classical information that can reliably be transmitted in this process is known as super dense coding capacity.
It has been shown that for noiseless
channels and unitary encoding, the capacity is given by
$C=\log d+S(\rho_b)-S(\rho)$ \cite{ ziman,hiroshima}.
Here, $\rho_b$ is Bob's
reduced density operator with $\rho_\textmd {b}=\mathrm{tr_a}\rho$, and
$S(\rho)=-\mathrm{\tr}(\rho \log \rho) $ is the von Neumann entropy.
Without the additional resource of entangled states, a
$d$-dimensional quantum state can be used to transmit the
information  $\log d$. Hence, quantum states  for which
$S(\rho_b)-S(\rho)>0 $, i.e., those which are more mixed locally
than globally, are the useful states for super dense coding.

A realistic quantum system usually suffers from unwanted
interactions with the outside world. 
Optical
 fibers and an unmodulated spin chain \cite{Bose-channel} are examples of such quantum channels
 which are suitable for long- and short-distance quantum communication, respectively.
Super dense coding in the situation when the quantum
states experience noise in the transmission channels was studied
in \cite{zahra-paper}. 
In \cite{zahra-paper} uncorrelated noise (i.e., 
memoryless channels) was discussed. For those cases (channels
and states) where the von Neumann entropy
 fulfills a specific condition, the super dense coding capacity was derived. Explicitly, for the two-dimensional uncorrelated
 depolarizing channel, it was shown that Alice and Bob do not win by sending classical information via a super dense
 coding protocol with unitary encoding if there is too much noise.

In this paper, memory effects along the transmission channel 
are taken into account. In this scenario the noisy channel acting on two
subsystems cannot be expressed as a product of two independent channels acting
on each subsystem separately. In particular, we
investigate the bipartite super dense coding scenario for a 
\emph{correlated} Pauli channel and
unitary or non-unitary encoding. Such kinds of channels were originally 
analyzed from the point of view of optimization of the classical information 
transmission \cite{mp,memory-quasi-chiara,cerf}.

The paper is organized as follows. In 
Sec. \ref{1-1-0} we review the Holevo bound as a
key concept in finding the super dense coding capacity. 
We discuss the mathematical definition of the
Holevo quantity in the presence of an arbitrary channel $\Lambda$.
Section \ref{1-3} is devoted to the super dense coding
capacity in the presence of a {correlated} Pauli channel 
and unitary encoding. We give examples of correlated
channels and initial states for which  the capacity is explicitly determined. Section \ref{non-unitary encoding} is dedicated to the {correlated} Pauli channel and  non-unitary encoding. We compare the capacities related to both unitary and non-unitary encoding and also discuss a case where non-unitary encoding has an advantage over unitary encoding.
Finally, in Sec. \ref{conc}, we summarize the main results. 

\section{capacity of super dense coding \label{1-1-0}}

The performance of a given composite state $\rho$ for super dense coding is 
usually quantified by the Holevo quantity, maximized over all possible encodings on Alice's side.
A theorem stated by Gordon \cite {Gordon} and Levitin  \cite{Levitin}, 
and proved by Holevo \cite {Holevo-chi-quantity}, states
that the amount of accessible classical information ($I_{acc}$)
contained in an ensemble $\{ \rho_i, p_i\}$ is upper bounded by
the so-called $\chi$-quantity $\chi(\{ \rho_i, p_i\})$, often also referred to as the  Holevo quantity. This upper
bound holds for any measurement that can be performed on
the system, and is given by
\begin{eqnarray}
I_{acc}\leq \chi(\{ \rho_i, p_i\})\equiv
S\left(\overline{\rho}\right)-\sum_i p_i S\left(\rho_i\right),
\label{chi-quantity1}
\end{eqnarray}
where $\overline{\rho}=\sum_i p_i \rho_i$ is the average ensemble
state and $ S(\eta)=-\mathrm{\tr}(\eta \log \eta)$ is the von Neumann
entropy of $\eta $. From the concavity of the von Neumann entropy
$S(\rho)$ it follows that the Holevo quantity is non-negative. 
The Holevo bound (\ref{chi-quantity1}) is achievable in
the asymptotic limit \cite{Schumacher-Westmoreland, Holevo-capacity}.

\subsection {Holevo quantity in the presence of noise \label{1-1-3}}
A quantum channel is a communication channel
which can transmit quantum information. Physically, a noisy
quantum channel is a communication channel that is affected by interaction
 with the environment. Mathematically, a noisy quantum channel can be described  as a completely positive trace
preserving (CPTP) map acting  on the quantum state that is transmitted. 
We consider
$\Lambda:\rho_i \rightarrow \Lambda(\rho_i)$ to be a CPTP map that
acts on the encoded  state $\rho_i= (W_i\otimes\id)\rho({W_i}^{\dagger}\otimes\id)$. For $\{\Lambda(p_i,\rho_i)\}$
being the ensemble that Bob receives, the Holevo quantity is given by
\begin{eqnarray}
\chi\{\Lambda(\rho_i,p_i)\}&=&S\left(\overline{\Lambda(\rho)}\right)-\sum_i
p_i S\left(\Lambda(\rho_i)\right)\nonumber\\
&=&\sum_{i}p_iS\left(\Lambda(\rho_i)\|\overline{\Lambda(\rho)}\right),
\label{Holevo}
\end{eqnarray}
where $\overline{\Lambda(\rho)}=\sum_i p_i \Lambda(\rho_i)$ is the
average state after transmission through the noisy channel and $ S(\rho \parallel \sigma)=\mathrm{\tr} \rho \left(\log
\rho - \log \sigma \right)$ is the relative entropy. The super dense coding capacity
$C$ for a given resource state $\rho$ and the noisy channel
$\Lambda$ is defined to be the maximum of the  Holevo quantity
$\chi{\{\Lambda(p_i,\rho_i)\}} $ with respect to the unitary operators $W_i$, chosen with the probabilities $p_i$, namely
\begin{eqnarray}
 C&=&\max_{\{W_i,p_i \}}(\chi{\{\Lambda(p_i,\rho_i)\}})\nonumber\\
&\equiv& \max_{\{W_i,p_i \}}
 \left(S\left(\overline{\Lambda(\rho)}\right)-\sum_i p_i S\left(\Lambda(\rho_i)\right)\right).
\end{eqnarray}
In the following, we will concentrate on the optimization 
of the Holevo quantity  in order to find the super dense coding
capacity.

\section{ Super dense coding via  correlated Pauli channels\label{1-3}}
We will now consider quantum channels with memory, where noise in consecutive 
uses of the channel is correlated. We specifically consider correlated Pauli 
channels \cite{mp,memory-quasi-chiara,cerf}, modelled as follows. 
Consider first a single Pauli channel, 
whose action on a $d$-dimensional density 
operator $\xi$ is given by
\begin{eqnarray}
\Lambda^{\textmd {P}}(\xi)=\sum_{m,n=0}^{d-1}q_{mn} V_{mn}\xi V_{mn}^\dagger\;,
\label{pauli-d-channel}
\end{eqnarray}
where $q_{mn}$ are probabilities (i.e., $q_{mn}\geq 0$ and
$\sum_{mn}q_{mn}=1$). The unitary \emph{displacement} operators $V_{mn}$ are 
defined as
\begin{eqnarray}
V_{mn}=\sum_{k=0}^{d-1}\exp \left({\frac{2i\pi kn}{d}}\right)\ket{k}\bra{k+m(\mathrm{mod}\,
{d})}\;.
\label{vmn}
\end{eqnarray}
The above operators satisfy $\mathrm{\tr} V_{mn} = d\delta_{m0} \delta_{n0} $ 
and $V_{mn} V_{mn}^\dagger=\id$, and commute up to a phase,
\begin{eqnarray}
 V_{mn} V_{\tilde{m}\tilde{n}}
=\exp\left({\frac{2i\pi(\tilde{n}m-n\tilde{m})}{d}}\right) V_{\tilde{m}\tilde{n}}V_{mn}.
\label{vv}
\end{eqnarray}
As the  operators $V_{mn}$ in  Eq. (\ref{pauli-d-channel}) are unitary, the
Pauli channel is unital, i.e., it preserves the identity. Now, let
$\Lambda^P_a$ and $\Lambda^P_\textmd {b}$ be two $d$-dimensional Pauli
channels (\ref{pauli-d-channel}) with the elements $\{q_{mn},V_{mn}\}$ and $\{q_{\tilde{m}\tilde{n}},V_{\tilde{m}\tilde{n}}\}$, respectively. Based on the elements of these two channels, a model of a {correlated} Pauli 
channel is defined as
\begin{eqnarray}
\Lambda^{\textmd P}_{\textmd {ab}}(\xi)=\sum_{m,n,\tilde{m},\tilde{n}=0}^{d-1}
 q_{mn\tilde{m}\tilde{n}} (V_{mn}\otimes V_{\tilde{m}\tilde{n}})\xi
  (V_{mn}^\dagger\otimes V_{\tilde{m}\tilde{n}}^\dagger),\nonumber\\
\label{k-Pauli-channels}
\end{eqnarray}
 where the probability $q_{mn\tilde{m}\tilde{n}}$ is given by
$q_{mn\tilde{m}\tilde{n}}=(1-\mu)q_{mn}q_{\tilde{m}\tilde{n}}+\mu q_{mn}\delta_{m,\tilde{m}}\delta_{n,\tilde{n}}$, and the parameter $\mu$ ($0\leq\mu\leq1$) 
quantifies  the correlation degree. For
 $\mu=0$ the two channels $\Lambda^P_a$ and $\Lambda^P_\textmd {b}$ are uncorrelated and act independently on Alice's and Bob's subsystems, respectively. For $\mu=1$, the global channel (\ref{k-Pauli-channels}) is called \emph{fully
  correlated} and for other values of $\mu$, different from 
\emph{zero} and \emph{one}, the global channel is partially correlated.

For a single sender, a single receiver, and a {correlated} Pauli
 channel as well as unitary and non-unitary encoding, we  derive two explicit
   expressions for the super dense coding capacity. We show that both unitary and non-unitary
   encoding problems reduce to the problem of finding a single CPTP map (in the case of unitary encoding this is a specific unitary transformation)
  that minimizes the output von Neumann entropy after its application
    and the action of the channel on the input state $\rho$. For the case of
   unitary encoding we find examples for the optimal unitary operator.

\subsection {Unitary encoding \label{unitary-correlated-Pauli}}
This subsection treats the optimization of the Holevo
quantity  for a {correlated} Pauli channel and unitary
encoding. We first introduce an upper bound on the Holevo quantity and
we then show that this upper bound is reachable and thus coincides with 
the super
dense coding capacity. This procedure is phrased in the following
Lemma.\newline \textbf{Lemma 1.} Let
\begin{eqnarray}
\chi=S\left( \overline{\Lambda^{\textmd P}_{\textmd {ab}}(\rho)}\right)-\sum_i p_i S
\left(\Lambda^{\textmd P}_{\textmd {ab}}\left(\rho_i\right)\right) \label{Holevo-un}
\end{eqnarray}
 be the Holevo quantity with  $
\rho_i=(W_i\otimes\id)\rho({W_i}^{\dagger}\otimes\id)$, the
average state $\overline{\Lambda^{\textmd P}_{\textmd {ab}}(\rho)}=\sum_i p_i
\Lambda^{\textmd P}_{\textmd {ab}}(\rho_i)$ and $\Lambda^{\textmd P}_{\textmd {ab}}$ the
{correlated} Pauli channel defined via Eq.
(\ref{k-Pauli-channels}).  Let $U_{\textmd{min}}$  be the unitary operator
that minimizes the von Neumann entropy after application of this
unitary operator and  the channel $\Lambda_{\textmd {ab}}^P$ to the initial
state $\rho$, i.e., $U_{\textmd{min}}$ minimizes the expression
 $S\left(\Lambda^{\textmd P}_{\textmd {ab}}( (U_{\textmd{min}}\otimes\id)\rho(U_{\textmd{min}}^\dagger\otimes\id))\right)$.
 Then the super dense coding capacity  $C^{\textmd{P}}_{\textmd{un}}$ is given by
\begin{eqnarray}
C^{\textmd{P}}_{\textmd{un}}&=&\log d +
S\left(\Lambda^\textmd{P}_\textmd {b}(\rho_\textmd {b})\right)\nonumber\\
&-&S\left(\Lambda^{\textmd P}_{\textmd {ab}}\left(
(U_{\textmd{min}}\otimes\id)\rho(U_{\textmd{min}}^\dagger\otimes\id)\right)\right),
\label{c-unitary-p}
\end{eqnarray}
where  $\rho_\textmd {b}=\mathrm{\tr_a} \rho$ and $\Lambda^\textmd{P}_\textmd {b}$ is the $d-$dimensional
Pauli channel (\ref{pauli-d-channel}) on Bob's subsystem. The subscript \textquotedblleft \textmd{un}\textquotedblright \hspace{0.2mm} refers to unitary encoding.\\
\textbf{Proof:} We start by introducing an upper bound on the
Holevo quantity (\ref{Holevo-un}). Since $U_{\textmd{min}}$ is a unitary
operator that leads to the minimum of the output von Neumann
entropy, for $\chi$ we have
\begin{eqnarray}
\chi
&\leq& S\left(
\overline{\Lambda^{\textmd P}_{\textmd {ab}}(\rho)}\right)-S\left(\Lambda^{\textmd P}_{\textmd {ab}}\left(
(U_{\textmd{min}}\otimes\id)\rho(U_{\textmd{min}}^\dagger \otimes\id)\right)\right).\nonumber
\end{eqnarray}
The von Neumann entropy is subadditive and the maximum entropy of
a $d$-dimensional system is $\log d $. Therefore we can write
\begin{eqnarray}
\chi &\leq&\log d +S\left(\mathrm{\tr_a}
\overline{\Lambda^{\textmd P}_{\textmd {ab}}(\rho)}\right)  \nonumber\\
&-&
S\left(\Lambda^{\textmd P}_{\textmd {ab}}\left(
(U_{\textmd{min}}\otimes\id)\rho(U_{\textmd{min}}^\dagger\otimes\id)\right)\right)\nonumber\\
 &=& \log d + S\left(\Lambda^\textmd{P}_\textmd {b}(\rho_\textmd {b})\right)\nonumber\\
&-&S\left(\Lambda^{\textmd P}_{\textmd {ab}}\left(
(U_{\textmd{min}}\otimes\id)\rho(U_{\textmd{min}}^\dagger\otimes\id)\right)\right),
\label{upperbound-unitary}
\end{eqnarray}
where we have used that $\mathrm{\tr_a} \overline{\Lambda^{\textmd P}_{\textmd {ab}}(\rho)}=\Lambda^\textmd{P}_\textmd {b}(\rho_\textmd {b})$. This
statement can be proved as follows. By using the definition of a {correlated} Pauli channel eq.  (\ref{k-Pauli-channels}) and of the average state, and noting that $W_i$ acts on Alice's side, we have
\begin{eqnarray}
\mathrm{\tr_a} \overline{\Lambda^{\textmd P}_{\textmd {ab}}(\rho)}&=& \sum_i p_i \sum_{m,n,\tilde{m},\tilde{n}=0}^{d-1}
 q_{mn\tilde{m}\tilde{n}} \mathrm{\tr_a} \Big[(V_{mn}\otimes V_{\tilde{m}\tilde{n}})\nonumber\\
&&(W_i\otimes\id)\rho({W_i}^{\dagger}\otimes\id) (V_{mn}^\dagger\otimes V_{\tilde{m}\tilde{n}}^\dagger)\Big]\nonumber\\
&=& \sum_i p_i \sum_{m,n,\tilde{m},\tilde{n}=0}^{d-1}  q_{mn\tilde{m}\tilde{n}}     V_{\tilde{m}\tilde{n}} \rho_\textmd {b}  V_{\tilde{m}\tilde{n}}^\dagger\nonumber\\
&=&\sum_{\tilde{m}\tilde{n}}q_{\tilde{m}\tilde{n}}V_{\tilde{m}\tilde{n}} 
\rho_\textmd {b}  
V_{\tilde{m}\tilde{n}}^\dagger=\Lambda^\textmd{P}_\textmd {b}(\rho_\textmd {b}),
\end{eqnarray}
which completes this part of the proof.
To show that the upper bound (\ref{upperbound-unitary}) is achievable, 
we consider
the ensemble $\{ \tilde{ p_i}=\frac{1}{d^2},\tilde{U_i}=V_iU_{\textmd{min}}\}$
 with $V_{i(=mn)}$ being the displacement operators of Eq. (\ref{vmn}). The Holevo quantity for
this ensemble is denoted by $\tilde{\chi}$ and is given by
\begin{eqnarray}
\tilde{\chi}&=& S \Big( \sum_i\frac{1}{d^2}\Lambda^{\textmd P}_{\textmd {ab}}\left(
(\tilde{U_i}\otimes\id)\rho({\tilde{U_i}}^{\dagger}\otimes\id)
 \right)\Big)\nonumber\\
&-&\sum_i \frac{1}{d^2}
S\left(\Lambda^{\textmd P}_{\textmd {ab}}\left(
(\tilde{U_i}\otimes\id)\rho({\tilde{U_i}}^{\dagger}\otimes\id) \right)\right).
\label{optiensemble-unitary}
\end{eqnarray}
 In \cite{hiroshima}, for an arbitrary bipartite state $\tau$, it was shown that $ \frac{1}{d^2}\sum_i (V_i\otimes\id) \tau ({V_i}^{\dagger}\otimes\id)=\frac{\id}{d} \otimes \mathrm{\mathrm{\tr_a}} \tau $.
By using this property, and noting that
$U_{\textmd{min}}$ acts only on Alice's side, we find that the argument in
the first term on the RHS of (\ref{optiensemble-unitary}) is given
by
\begin{eqnarray}
&&\sum_i\frac{1}{d^2}\Lambda^{\textmd P}_{\textmd {ab}}\left((\tilde{U_i}
\otimes\id)\rho({\tilde{U_i}}^{\dagger} \otimes\id)\right)\nonumber\\
&=&\Lambda^{\textmd P}_{\textmd {ab}}\sum_i\frac{1}{d^2}\left( (V_iU_{\textmd{min}}
\otimes\id)\rho(U_{\textmd{min}}^\dagger {V_i}^{\dagger}\otimes\id)\right)\nonumber\\
&=&\Lambda^{\textmd P}_{\textmd {ab}}(\frac{\id}{d}\otimes \rho_\textmd {b} ) =\frac{\id}{d}
\otimes\Lambda^{\textmd P}_\textmd {b}({\rho_\textmd {b}}). \label{average-state1}
\end{eqnarray}
Furthermore, the second term on the RHS of Eq. (\ref{optiensemble-unitary}) can be expressed in terms of the
unitary operator $U_{\textmd{min}}$ and the channel. By inserting the action of
the {correlated} Pauli  channel, using Eq. (\ref{vv}), from
which  follows that $V_{i(=jk)}$ and $V_{mn}$ commute up to a
phase, and since the von Neumann entropy is invariant under unitary
transformation, we can write
\begin{eqnarray}
&&\sum_i \frac{1}{d^2}  S\left(\Lambda^{\textmd P}_{\textmd {ab}}( \tilde{U_i}
\otimes\id )\rho({\tilde{U_i}}^{\dagger} \otimes\id )
\right)\nonumber\\
&=&\frac{1}{d^2} \sum_{kj} S\Big((V_{jk}\otimes \id )
\Big[\sum_{m,n,\tilde{m},\tilde{n}} q_{mn\tilde{m}\tilde{n}} \left(V_{mn}\otimes V_{\tilde{m}\tilde{n}}\right)
 \nonumber\\
&& \left( U_{\textmd{min}} \otimes\id\right) \rho
(U_{\textmd{min}}^\dagger \otimes\id)(V_{mn}^\dagger
\otimes V_{\tilde{m}\tilde{n}}^\dagger )\Big]
 (V_{jk}^\dagger\otimes \id)\Big)\nonumber\\
&=& S\left(\Lambda^{\textmd P}_{\textmd {ab}}\left( \left( U_{\textmd{min}} \otimes\id\right)
\rho (U_{\textmd{min}}^\dagger \otimes\id) \right)\right).
\label{average-entropy1}
\end{eqnarray}
Inserting Eqs. (\ref{average-state1}) and (\ref{average-entropy1}) into
Eq. (\ref{optiensemble-unitary}), one finds that the Holevo quantity
$\tilde{\chi} $ is equal to  the upper bound given in
Eq. (\ref{upperbound-unitary}) and consequently, this is the super dense
coding capacity. \hfill {$\Box$}

By Lemma 1, we  proved that, in order to determine the super
dense coding capacity, it is enough to find an optimal $U_{\textmd{min}}$ that minimizes the
channel output von Neumann entropy $
S\left(\Lambda_{\textmd {ab}}^\textmd {P}\left((U_{\textmd{min}}\otimes\id)\rho
(U_{\textmd{min}}^\dagger
 \otimes\id)\right)\right)$. In the next two sections we give examples
 of channels and
initial states for which $U_{\textmd{min}}$ can explicitly be determined.

\subsection{Correlated quasi-classical channel \label{quasi}}

 A $d-$dimensional quasi-classical depolarizing
channel (or simply quasi-classical channel) is a particular form of
a $d-$dimensional Pauli channel \cite{memory-quasi-chiara,cerf}, 
as given in Eq. (\ref{pauli-d-channel}). For this channel, the
probabilities of  the { displacement }operators $V_{mn}$ are equal for $m=0$ 
and any phase shift labeled by $n$,
and they differ from the rest of the probabilities
which are also equal, i.e.,
\begin{eqnarray}
q_{mn}=
\begin{cases}
  \frac{1-p}{d},  & m=0 \\
  \frac{p}{d(d-1)}, & \mbox{otherwise}.
\end{cases}
\label{qmn-quasiclassica}
\end{eqnarray}
 The quasi-classical  channel is characterized by a single probability
parameter $ 0 \leq p\leq 1$. With probability $p$,
a {displacement} occurs and with probability $1-p$, no
{displacement} occurs to the quantum signal. Like in the
classical case, $p$ can also be seen as the amount of noise in a
channel.

In the following we will consider as a resource state a Werner state 
$\rho_{\textmd w}=  \eta
\ket{\varPhi^+}\bra{\varPhi^+}+\frac{1-\eta}{4}\id $ with
$\ket{\varPhi^{+}}=\frac{1}
{\sqrt{2}}\left(\ket{00}+\ket{11}\right)$. In the presence of
a {correlated} quasi-classical channel, as defined via Eqs. (\ref{k-Pauli-channels}) and (\ref{qmn-quasiclassica}), we find $U_{\textmd{min}}$.  
Thus the dimension is $d=2$. For two-dimensional systems the
{displacement} operators, defined in Eq. (\ref{vmn}), 
are either the identity or the Pauli operators, i.e.,
\begin{eqnarray}
\sigma_0=\begin{pmatrix}
  1 & 0 \\
  0 & 1
\end{pmatrix},\hspace{1.5mm}
\sigma_1=\begin{pmatrix}
  0& 1 \\
  1& 0
\end{pmatrix},\hspace{1.5mm}\nonumber\\
\sigma_2=\begin{pmatrix}
  0 & -i \\
  i & 0
\end{pmatrix},\hspace{1.5mm}
\sigma_3=\begin{pmatrix}
  1 & 0 \\
  0 & -1
\end{pmatrix}.
\label{pauli-def2}
\end{eqnarray}
The {correlated} quasi-classical  channel is
in this case
\begin{eqnarray}
\Lambda^Q_{\textmd {ab}} (\xi)= \sum_{m,n} q_{mn} \sigma_m\otimes\sigma_n (\xi)\sigma_m\otimes\sigma_n,
\end{eqnarray}
where $q_{mn}=  (1-\mu) q_m q_n+ \mu q_n\delta_{mn} $ with
$q_0=q_3=\frac{1-p}{2}$ and $q_1=q_2=\frac{p}{2}$. In order to
find $U_{\textmd{min}}$, we start with the most general $2\times 2$ unitary
operator $U$
\begin{eqnarray}
U=\begin{pmatrix}
  a & b \\
  -b^\ast & a^\ast
\label{general-unitary-operator}
\end{pmatrix},
\end{eqnarray}
where $a$ and $b$ are complex variables which satisfy $\lvert a
\lvert^2+\lvert b \lvert^2=1 $. The output of a {correlated}
quasi-classical channel for an arbitrary input state $\rho$ is invariant
under the unitary transformation $\sigma_3\otimes\sigma_3$ \cite{memory-quasi-chiara}, i.e.,
\begin{eqnarray}
\Lambda^Q_{\textmd {ab}} (\rho)=\Lambda^Q_{\textmd {ab}} \big((\sigma_3
\otimes\sigma_3 )\rho(\sigma_3\otimes\sigma_3)\big).
\label{invariant-qusi-sima3-2}
\end{eqnarray}
By the above property (\ref{invariant-qusi-sima3-2}), the channel outputs  for the
input states  $(U\otimes\id)\rho_{\textmd w} (U^\dagger\otimes\id)$ and
$(\sigma_3 U\otimes\sigma_3
)\rho_{\textmd w}(U^\dagger\sigma_3\otimes\sigma_3)$ are equal and therefore
we can conveniently use the average of both states instead of only $
(U\otimes\id)\rho_{\textmd w} (U^\dagger\otimes\id)$. This average is
given by
\begin{eqnarray}
&& \frac{1}{2}(U\otimes\id)\rho_{\textmd w} (U^\dagger\otimes\id)+\frac{1}{2}(\sigma_3 U\otimes\sigma_3 )\rho_{\textmd w}(U^\dagger\sigma_3\otimes\sigma_3)\nonumber\\
&=&\begin{pmatrix}
  \eta\frac{ a a^\ast}{2}+\frac{1-\eta}{4} & 0 & 0  &\eta  \frac{a^2}{2} \\
   0 & \eta \frac{b b^\ast}{2}+\frac{1-\eta}{4} & \eta \frac{- b^2}{2} & 0 \\
   0 &\eta  \frac{-(b^\ast)^ 2}{2} & \eta \frac{b b^\ast}{2}+\frac{1-\eta}{4}& 0\\
   \eta \frac{(a^\ast)^2}{2} & 0 & 0 & \eta \frac{a a^\ast}{2}+\frac{1-\eta}{4}
\end{pmatrix}\nonumber\\
&=&\lvert a\lvert^2  \Big(
\eta\ket{\varPhi_1}\bra{\varPhi_1}+\frac{1-\eta}{4}\id \Big)
+\lvert b \lvert^2
\Big(\eta\ket{\varPhi_2}\bra{\varPhi_2}+\frac{1-\eta}{4}\id\Big) ,\nonumber\\
\label{matrix-werner}
\end{eqnarray}
with $\ket{\varPhi_1}$ and $\ket{\varPhi_2}$ being
\begin{subequations}
\begin{eqnarray}
&&\ket{\varPhi_1}=\frac{1}{\sqrt{2}}\Big(\frac{a}{\lvert a
\lvert}\ket{00}+ \frac{a^\ast}{\lvert a \lvert}\ket{11}\Big),
\label{phi1}\\
&&\ket{\varPhi_2}=\frac{1}{\sqrt{2}}\Big(\frac{b}{\lvert b \lvert}\ket{01}- \frac{b^\ast}{\lvert b \lvert}\ket{10}\Big).
\label{phi2}
\end{eqnarray}
\end{subequations}
After  applying the quasi-classical channel, using Eqs. 
(\ref{invariant-qusi-sima3-2}) and (\ref{matrix-werner}), and the concavity of the von Neumann entropy, we
find the following lower bound
\begin{eqnarray}
&&S\Big(\Lambda^Q_{\textmd {ab}} \left((U\otimes\id)\rho_{\textmd w}(U^\dagger\otimes\id) \right)\Big)\nonumber\\
&\geq& \lvert a\lvert^2 S\Big( \Lambda^Q_{\textmd {ab}} \Big( \eta\ket{\varPhi_1}\bra{\varPhi_1}+\frac{1-\eta}{4}\id\Big)\Big)\nonumber\\
&+&\lvert b \lvert^2 S\Big(\Lambda^Q_{\textmd {ab}} \Big(\eta\ket{\varPhi_2}\bra{\varPhi_2}+\frac{1-\eta}{4}\id \Big)\Big)\nonumber\\
&\geq&  S\Big( \Lambda^Q_{\textmd {ab}} \Big(
\eta\ket{\varPhi^+}\bra{\varPhi^+}+\frac{1-\eta}{4}\id
\Big)\Big). \label{qclassicalwerner}
\end{eqnarray}
In the last line we have used that both $S\Big( \Lambda^Q_{\textmd {ab}}
\left(
\eta\ket{\varPhi_{1,2}}\bra{\varPhi_{1,2}}+\frac{1-\eta}{4}\id
\right)\Big)$ are lower bounded by $S\Big( \Lambda^Q_{\textmd {ab}} \left(
\eta\ket{\varPhi^+}\bra{\varPhi^+}+\frac{1-\eta}{4}\id
\right)\Big)$.
The proof for this
statement is as follows. We can rewrite $\ket{\varPhi_1}$ (a
similar argument holds for $\ket{\varPhi_2}$) up to a global phase as
\begin{eqnarray}
\ket{\varPhi_1}= \frac{1}{\sqrt{2}}\left(\ket{00} + \exp (i\phi) \ket{11} \right).
\label{phi-zero}
\end{eqnarray}
After applying the  {correlated} quasi-classical channel
$\Lambda^Q_{\textmd {ab}}$ to the state $ \eta\ket{\varPhi_{1}}\bra{\varPhi_{1}}+\frac{1-\eta}{4}\id$, we arrive at
\begin{eqnarray}
&&\Lambda^Q_{\textmd {ab}}\left(
\eta\ket{\varPhi_{1}}\bra{\varPhi_{1}}+\frac{1-\eta}{4}\id
\right)= \frac{1-\eta}{4}\id\otimes\id\nonumber\\
&+& \frac{\eta}{4}
\Big[ \left( \mu+(1-\mu)(1-2p)^2   \right) \sigma_3\otimes \sigma_3  \nonumber\\
&+&\mu (1-2p)\sin \phi(\sigma_1\otimes \sigma_2+\sigma_2\otimes \sigma_1)\nonumber\\
&+&\mu \cos \phi (\sigma_1\otimes \sigma_1-\sigma_2\otimes \sigma_2)+\id\otimes\id \Big].
\label{quasiclasical-varPhi-1}
\end{eqnarray}
The von Neumann entropy of a quantum state is defined via its
eigenvalues. The eigenvalues of Eq. (\ref{quasiclasical-varPhi-1}) are
\begin{eqnarray}
\nu_{1,2}&=& \eta(1-\mu) p (1 - p)+\frac{1-\eta}{4}\;, \nonumber\\
\nu_{3,4}&=& \frac{\eta}{2}\Big( 1- 2(1-\mu) p (1 - p)   \nonumber\\
&\pm&
\sqrt{\mu^2 (1-4 p(1-p ) {\sin^2 \phi}  )}\Big)+ \frac{1-\eta}{4}.
\label{eigenvaleus-quasi-phi}
\end{eqnarray}
To minimize the von Neumann entropy $S\left[\Lambda^Q_{\textmd {ab}} \left(
\eta\ket{\varPhi_1}\bra{\varPhi_1}+\frac{1-\eta}{4}\id \right)
\right]=-\sum_i \nu_i \log \nu_i $,
the eigenvalues should diverge as much as possible with respect to
the parameter $\phi$. The eigenvalues $\nu_{1,2}$ are
independent of $\phi$. Thus, the von Neumann entropy is minimized when we
maximize $\nu_{3}$ while we minimize $\nu_{4}$. This is the case for $\phi=0$ and it leads $\ket{\varPhi_{1,2}}$ to be the Bell
state $\ket{\varPhi^{+}}$ which proves the above statement, namely
\begin{eqnarray}
&&S\Big(\Lambda^Q_{\textmd {ab}} \Big(\eta\ket{\varPhi_{1,2}}\bra{\varPhi_{1,2}}+\frac{1-\eta}{4}\id \Big)\Big)\nonumber\\
&\geq&  S\Big( \Lambda^Q_{\textmd {ab}} \Big(
\eta\ket{\varPhi^+}\bra{\varPhi^+}+\frac{1-\eta}{4}\id
\Big)\Big).
\end{eqnarray}
\hfill{$\Box$}

The lower bound on the von Neumann entropy (\ref{qclassicalwerner}) is
 reachable: It is not difficult to see that the variables $a=1$ and $b=0$, which correspond  to $U$ being the identity operator.
We have thus found a unitary operator
 which minimizes the output entropy. Therefore, the super dense coding capacity for a Werner state in a
{correlated} quasi-classical channel,
according to Eq. (\ref{c-unitary-p}), is given by
\begin{eqnarray}
C_{\textmd{un}}^{\textmd{Q,w}}=2-S\left(\Lambda^Q_{\textmd {ab}}\left(\rho_{\textmd w} \right)\right).
\label{werne-quasiclassical}
\end{eqnarray}

For $\eta=1$, the Werner state $\rho_{\textmd w}$ reduces to a Bell state $\ket{\varPhi^+}$. Therefore, the super dense coding capacity, according to Eq. (\ref{werne-quasiclassical}), for a Bell state and in the presence of a {correlated}
quasi-classical channel, is given by
\begin{eqnarray}
C_{\textmd{un}}^{\textmd {Q,B}}=2-S\left(\Lambda^Q_{\textmd {ab}}\left( \ket{\varPhi^+}\bra{\varPhi^+}\right)\right).
\label{C-un-Bell}
 \end{eqnarray}

In Fig. 1 and Fig. 2, we visualize the super dense coding capacity for the {correlated}  quasi-classical channel as a function of the parameters $\mu$, $\eta$ and $p$  [Eq.(\ref{werne-quasiclassical})]. In Fig. 1, we consider a Bell state, i.e., $\eta=1$, as a function of the noise parameter $p$ and the
correlation degree $\mu$.
In Fig. 2, the noise parameter is fixed to $p=0.05$ and we vary $\mu$ and the parameter $\eta$ characterising the Werner state.

\begin{center}
\begin{tabular}{c  }
\includegraphics[width=9 cm]{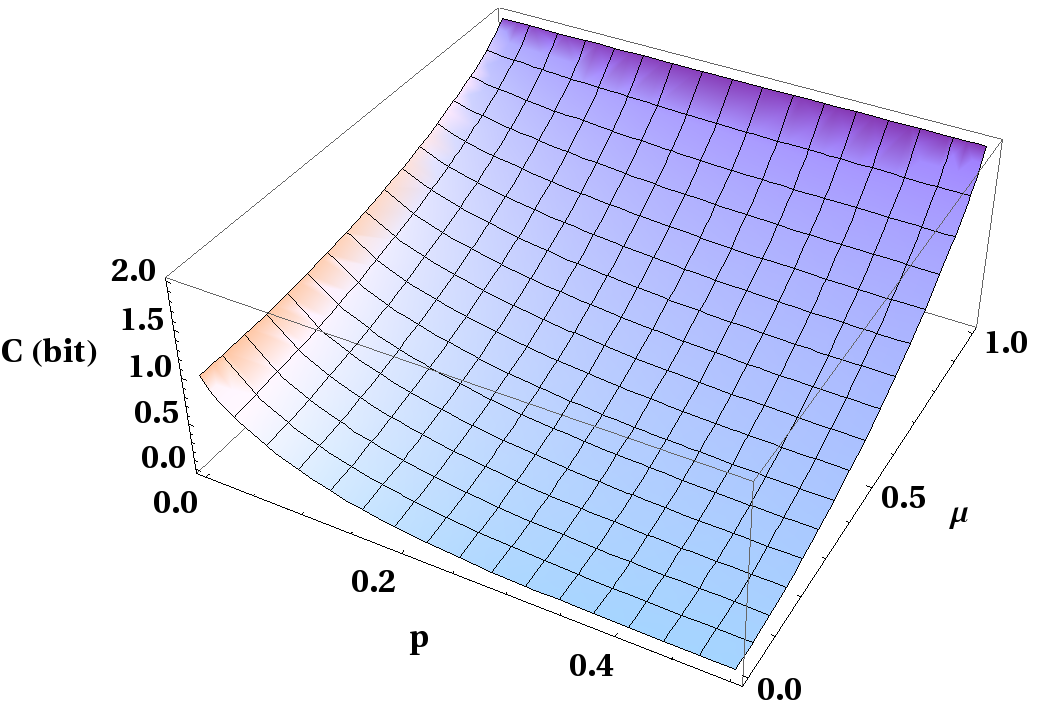}  \\
\end{tabular}
\end{center}

\vspace{0.5cm}

Fig. 1. (Color online)  The super dense coding capacity for a  correlated quasi-classical channel  and a Bell state ($\eta=1$ ), as a function of the noise parameter $p$ and the correlation degree $\mu$.

\vspace{0.5cm}

\begin{center}
\begin{tabular}{c  }
\includegraphics[width=9cm]{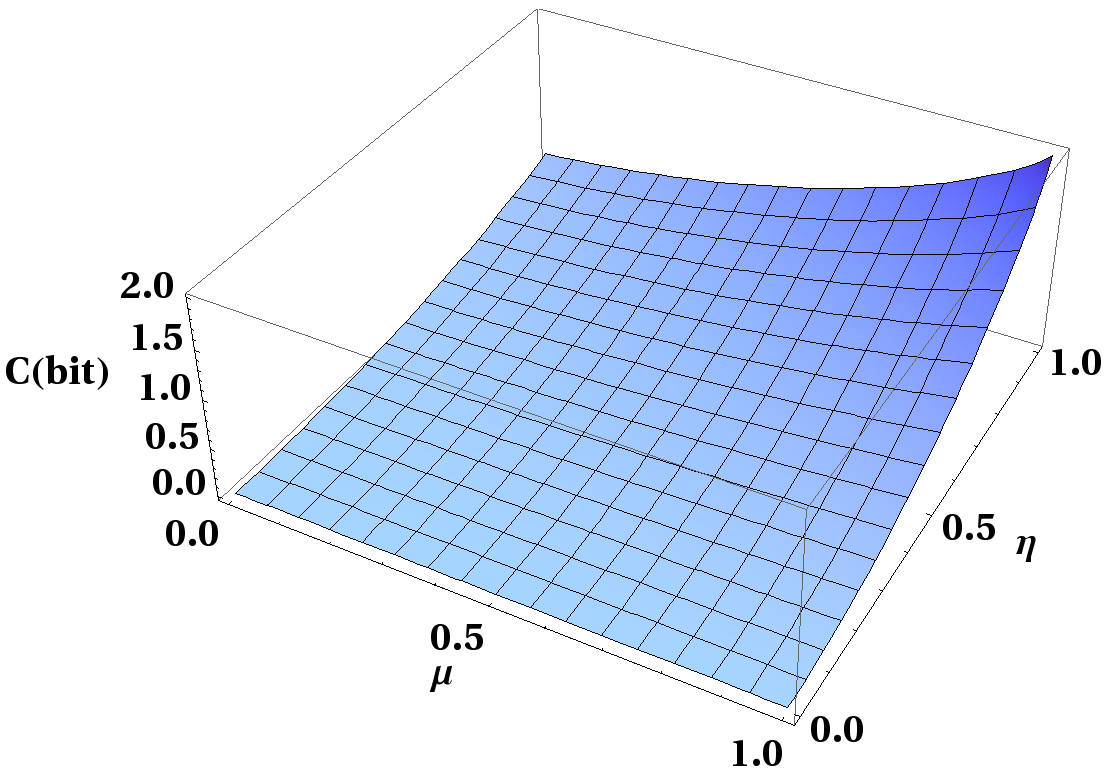}  \\
\end{tabular}
\end{center}

Fig. 2. (Color online) The super dense coding capacity for a correlated quasi-classical channel and a Werner state, as a function of the correlation degree $\mu$ and the  parameter $\eta$. The noise parameter is $p=0.05$.

\subsection{Fully correlated Pauli channel \label{fully-Pauli}}
In this section we give another example for which we determined
$U_{\textmd{min}}$. That is the case of a {fully correlated} Pauli
channel and a Werner state. As mentioned above, a
 {fully correlated} Pauli channel is a special form
of a {correlated} Pauli channel (\ref{k-Pauli-channels}) when
$\mu=1$.  For $d=2$, it is given by
\begin{eqnarray}
\Lambda^{\textmd f}_{\textmd {ab}}(\xi)
&=&\sum_{m}q_m(\sigma_m\otimes \sigma_m)(\xi)(\sigma_m\otimes \sigma_m),
\label{fully-pauli}
\end{eqnarray}
where $\sum_{m}q_m =1$ and $\sigma_m$ are either the identity or the Pauli
operators.

We again consider the Werner states $\rho_{\textmd w}=  \eta
\ket{\varPhi^+}\bra{\varPhi^+}+\frac{1-\eta}{4}\id $ as resource states.
For a {fully correlated} Pauli channel (\ref{fully-pauli}) we determine the operation $U_{\textmd{min}}$. To do so, we derive a
lower bound on $S\left(\Lambda^{\textmd f}_{\textmd {ab}}\left((U\otimes\id)\rho_{\textmd w}
(U^\dagger\otimes\id)\right)\right)$ where $U$ is an arbitrary
unitary operator. By using the concavity of the von Neumann
entropy and also by using the invariance of the von Neumann
entropy under unitary transformations, the lower bound on
$S\left(\Lambda^{\textmd f}_{\textmd {ab}}\left((U\otimes\id)\rho_{\textmd w}
(U^\dagger\otimes\id)\right)\right)$ takes the form
\begin{eqnarray}
&&S\left(\Lambda^\textmd{f}_{\textmd {ab}}\left((U\otimes\id)\rho_\textmd{w}
(U^\dagger\otimes\id)\right)\right)\nonumber\\
&=&S\Big(\sum_{m}q_m (\sigma_m\otimes \sigma_m)(U\otimes\id)  \nonumber\\
&&(\eta \rho^+ +\frac{1-\eta}{4}\id)(U^\dagger\otimes\id)(\sigma_m\otimes\sigma_m)\Big)\nonumber\\
&\geq&  S\Big( \eta \rho^+ + \frac{1-\eta}{4}\id  \Big)\;,
\label{fully-werner}
\end{eqnarray}
where we use the notation 
$\rho^+= \ket{\varPhi^+} \bra{\varPhi^+}$.

By using the invariance of a Bell state under the action of a
{fully correlated} Pauli channel, i.e., $\Lambda_{\textmd {ab}}^\textmd{f}
(\rho^+)=\rho^+$, it follows that the lower bound
(\ref{fully-werner}) is reachable by
 the identity operator. Then
$U_{\textmd{min}}=\id$ and the super dense coding capacity, according to
(\ref{c-unitary-p}), is given by
\begin{eqnarray}
C_{\textmd{un}}^{\textmd{f,w}}=2-S\left(\Lambda^{\textmd f}_{\textmd {ab}}\left(\rho_\textmd{w} \right)\right).
\label{c-fully-werner}
\end{eqnarray}

The Werner state $\rho_{\textmd w}$ reduces to a Bell state $\rho^+$
for $\eta=1$. Since the Bell state is invariant under the action of a
{fully correlated} Pauli channel, its  von Neumann entropy
   $S\left(\Lambda_{\textmd {ab}}^\textmd{f}\left(\rho^+ \right)\right)$ is zero. Therefore, using Eq. (\ref{c-fully-werner}), the super dense coding capacity for a shared Bell state and a {fully correlated} Pauli channel (\ref{fully-pauli}), is {two} bits. It is the
     maximum information transfer for $d=2$. This shows that no information at all is lost to the
environment and this class of channels behaves like a noiseless one. This behavior corresponds to the results of \cite{mp,memory-quasi-chiara}.

\section{Non-unitary encoding \label{non-unitary encoding}}
So far, we have assumed that the encoding in the super dense coding protocol
is unitary. The super
 dense coding protocol with non-unitary encoding for noiseless channels has been
 discussed by 
M. Horodecki and Piani \cite{CPTP}, M. Horodecki et al. \cite{unitaryoptimal1},
and Winter
\cite{unitaryoptimal2}. In this section we consider the possibility
of performing non-unitary encoding in the presence of a
{correlated} Pauli channel. Let us consider $\Gamma_i$ to be
a completely positive trace preserving (CPTP) map. Alice applies
 the map $\Gamma_i$ on her side of the shared state $\rho$, thereby
 encoding $\rho$ as $ \rho_i= [\Gamma_i \otimes \id]
 (\rho):=\Gamma_i(\rho)$. The rest of the scheme is similar to the
 case of unitary encoding. Alice sends the encoded state
 $ \rho_i=\Gamma_i(\rho)$ with the probability $p_i$ to Bob
 through the {correlated} Pauli channel $\Lambda_{\textmd {ab}}^P $.
 Now, the question is: which ensemble of CPTP maps achieves  the super
 dense coding capacity? In other words, what is the optimum  Holevo quantity with respect to
  the encoding $\Gamma_i$ and $p_i$?
To answer this question, first we give the definition for the
super dense coding capacity with a {correlated} Pauli channel
and non-unitary encoding:
\begin{eqnarray}
C &=&\max_{\{\Gamma_i , p_i\}}
\Big[S\Big( \sum_i p_i
\Lambda^{\textmd P}_{\textmd {ab}}\left(\Gamma_i(\rho)\right)\Big)\nonumber\\
&-& \sum_i p_i S
\Big(\Lambda^{\textmd P}_{\textmd {ab}}\left(\Gamma_i(\rho)\right)\Big)\Big],
\label{holevo-bound-non-unitary}
\end{eqnarray}
where $\Lambda^{\textmd P}_{\textmd {ab}}(\rho)$ is defined via
(\ref{k-Pauli-channels}). Similar to the unitary encoding case in
section \ref{unitary-correlated-Pauli}, we find an upper bound on
the Holevo quantity (\ref{holevo-bound-non-unitary}) and then we
show that this upper bound is reachable by a pre-processing before
unitary encoding. The above statement will be expressed in the
following Lemma.

\vspace{2mm}
\noindent \textbf{Lemma 2.} Let $\chi$ be
the Holevo quantity (\ref{holevo-bound-non-unitary}), and let
$\Gamma_ {\textmd{min}}(\rho):=[\Gamma_{\textmd{min}} \otimes \id](\rho)$  be the
map that minimizes the von Neumann entropy  after application of
this map and  the {correlated} Pauli channel $\Lambda^{\textmd P}_{\textmd {ab}}$ to the initial state
$\rho$, i.e., $\Gamma_{\textmd{min}}$ minimizes the expression
 $S\left(\Lambda^{\textmd P}_{\textmd {ab}}(\Gamma_{\textmd{min}} (\rho))\right)$.
 Then the super dense coding capacity  is given by
\begin{eqnarray}
 C^P_{\textmd {non-un}}=\log d + S\Big(\Lambda^\textmd {P}_\textmd{b} (\rho_\textmd{b})\Big)-
    S\Big(\Lambda^{\textmd P}_{\textmd {ab}}\big(\Gamma_{\textmd{min}}(\rho)\big)\Big).\nonumber\\
\label{capacity-non-unitary}
\end{eqnarray}
where $\rho_\textmd {b}=\mathrm{tr_a} \rho$ and $\Lambda^\textmd{P}_\textmd {b}$ is the $d-$dimensional
Pauli channel (\ref{pauli-d-channel}) on Bob's subsystem.

\vspace{2mm}
 \noindent \textbf{Proof:} The von Neumann entropy is sub-additive
and the maximum entropy of a $d$-dimensional system is $\log d $, and since $\Gamma_{\textmd{min}}$ is a map that leads to
the minimum of the entropy after applying it and the channel to
the initial state $\rho$,  we have the upper bound
\begin{eqnarray}
\chi
&\leq& S\left( \sum_i p_i
\Lambda^{\textmd P}_{\textmd {ab}}\left(\Gamma_i(\rho)\right)\right)- S
\left(\Lambda^{\textmd P}_{\textmd {ab}}\left(\Gamma_{\textmd{min}} (\rho)\right)\right).\nonumber\\
&\leq& \log d + S\left(\mathrm{\tr_a} \left(\sum_i p_i
\Lambda^{\textmd P}_{\textmd {ab}}\left(\Gamma_i(\rho)\right)\right)\right) - S
\left(\Lambda^{\textmd P}_{\textmd {ab}}\left(\Gamma_{\textmd{min}} (\rho)\right)\right).
\nonumber
\end{eqnarray}
By using $\mathrm{\tr_a} \sum_i p_i
\Lambda^{\textmd P}_{\textmd {ab}}\big(\Gamma_i(\rho)\big)=\Lambda^\textmd{P}_\textmd {b}(\rho_\textmd {b})$, we
find the upper bound
\begin{eqnarray}
\chi &\leq& \log d + S\Big(\Lambda^\textmd {P}_\textmd {b}(\rho_\textmd {b})\Big)-S \left(\Lambda^{\textmd P}_{\textmd {ab}}\left(\Gamma_{\textmd{min}}(\rho)\right)\right)
.\label{upperbound}
\end{eqnarray}
We now show that the ensemble $\{{\tilde{
p_i},\tilde{\Gamma}_i(\rho)}\}$ with $ \tilde{p_i}=\frac{1}{d^2}$
and $\tilde{\Gamma}_i(\rho)=(V_i \otimes\id) \Gamma_{\textmd{min}} (\rho)(V_i^\dagger \otimes\id)$ where $V_{i(=mn)}$ is defined in (\ref{vmn}),  reaches the upper bound (\ref{upperbound}). In other words, the optimal encoding consists of a fixed pre-processing
with $\Gamma_{\textmd{min}}$ and a subsequent unitary encoding. This is
analogous to the case of noiseless channels and uncorrelated Pauli channels \cite{CPTP,zahra-paper}. Below we prove the above
claim.

The Holevo quantity of the ensemble $\{{\tilde{
p_i},\tilde{\Gamma}_i(\rho)}\}$ is
\begin{eqnarray}
\tilde{\chi}= S \Big( \sum_i\frac{1}{d^2}\Lambda^{\textmd P}_{\textmd {ab}}\left(\tilde{\Gamma}_i(\rho)\right)\Big)-\sum_i \frac{1}{d^2}  S\left(\Lambda^{\textmd P}_{\textmd {ab}}\left(\tilde{\Gamma}_i(\rho)\right)\right).\nonumber\\
\label{optiensemble}
\end{eqnarray}
With an argument similar to the case of unitary encoding, the first term on the RHS of
(\ref{optiensemble}) is given by
\begin{eqnarray}
\sum_i\frac{1}{d^2}\Lambda^{\textmd P}_{\textmd {ab}}\left(\tilde{\Gamma}_i(\rho)\right)
=\frac{\id}{d} \otimes\Lambda^\textmd {P}_\textmd {b}({\rho_\textmd {b}}).
\label{average-state}
\end{eqnarray}
Furthermore, for the second term on the RHS of (\ref{optiensemble}) we have
\begin{eqnarray}
&&\sum_i \frac{1}{d^2}  S\left(\Lambda^{\textmd P}_{\textmd {ab}}\left(\tilde{\Gamma}_i(\rho)\right)\right)=S\left(\Lambda^{\textmd P}_{\textmd {ab}}\left({\Gamma_{\textmd{min}}}(\rho)\right)\right).
\label{average-entropy}
\end{eqnarray}
Inserting Eqs. (\ref{average-state}) and (\ref{average-entropy}) into
Eq. (\ref{optiensemble}), one finds that the Holevo quantity
$\tilde{\chi} $ is equal to  the upper bound given in
Eq. (\ref{upperbound}). Consequently, the super dense coding capacity
with non-unitary encoding is determined  by
Eq. (\ref{capacity-non-unitary}). \hfill{$\Box$}

A comparison of Eqs. (\ref{capacity-non-unitary}) and (\ref{c-unitary-p}) shows
that applying an appropriate pre-processing
$\Gamma_{\textmd{min}}$ on the initial state $\rho$ before the unitary
encoding $\{V_i\}$  may increase the super dense coding capacity,
with respect to only using unitary encoding for the case of a {correlated} Pauli channel. However, for some examples, no  better encoding than unitary encoding is possible. For instance, since \textit{two} bits  is the highest super dense coding capacity for $d=2$, our results derived in  Sec. \ref{fully-Pauli} for {fully
correlated} Pauli channel and the  Bell state provide  an example
where no pre-processing can improve the capacity. However,  examples exist for which non-unitary pre-processing is useful to increase the super dense coding capacity. In the next section we provide an explicit example.
\subsection{ Pre-processing can improve  capacity \label{Pre-processing improves}}

 Here, we show that for a two-dimensional Bell state in the presence of a 
{correlated} quasi-classical channel, a non-unitary pre-processing $\Gamma$, which is not necessarily $\Gamma_{min}$, can improve the super dense coding capacity. To show this claim, consider 
the completely positive trace preserving pre-processing $\Gamma$, with the Kraus operators $E_1=\ket{0}\bra{1}$ and $E_2=\ket{0}\bra{0}$. 
Alice applies $\Gamma$ on her side of the Bell state 
$\rho^+=\ket{\varPhi^+}\bra{\varPhi^+}$ and transforms the Bell state to  $\Gamma (\rho^+) =\ket{0}\bra{0} \otimes \frac{\id}{2}$. Therefore, according to Eq. (\ref{capacity-non-unitary}), for a {correlated} quasi-classical channel, a Bell state, and a pre-processing $\Gamma$, the amount of information that is transmitted by this process is given by
\begin{eqnarray}
C^\textmd{{Q,B}}_{\Gamma}=1+p \log p + (1-p) \log (1-p),
\label{non-unitary-qusi-Bell}
\end{eqnarray}
where $p$ is the noise parameter for a quasi-classical channel (\ref{qmn-quasiclassica}). Since $\Gamma$ is not necessarily the optimal pre-processing, $ C^\textmd{{Q,B}}_{\Gamma}$ is not also necessarily the capacity. 
We name (\ref{non-unitary-qusi-Bell}) the \emph{transferred information}. 
We now compare the {transferred information} (\ref{non-unitary-qusi-Bell}) with the capacity (\ref{C-un-Bell}) which is achieved by applying only unitary 
encoding. 
In the range of $0.3 \leqslant \mu \leqslant 1$
we find that the capacity $C^{Q,B}_{\textmd {un}}$ is always higher than 
the  {transferred information} $C^{Q,B}_{\Gamma}$, 
i.e., $C^{Q,B}_{\Gamma}\textless C^{Q,B}_{\textmd {un}}$. 
Therefore, in this range, the chosen pre-processing $\Gamma$ does not improve the capacity. 
In the range of $0\leqslant \mu \textless 0.3$, the capacity with unitary encoding (\ref{C-un-Bell}) and the {transferred information} with the pre-processing $\Gamma$ (\ref{non-unitary-qusi-Bell}) coincide for $\mu=\tilde{\mu}(p)$, shown as the  dashed red   curve in Fig. 3. 
Note that $\tilde{\mu}(p)$ corresponds to the 
$\textmd {Root}[C^{Q,B}_{\textmd {un}}-C^{Q,B}_{\Gamma}]$.
The function $\tilde{\mu}(p)$ is invariant under the simultaneous exchange  $p \leftrightarrow 1-p$ since both functions $C^{Q,B}_{\textmd {un}}$ and $C^{Q,B}_{\Gamma}$ are symmetric under the exchange $p \leftrightarrow 1-p$. Our results show that for $\mu\textless\tilde{\mu}(p)$, the  blue (light gray)  area in Fig. 3, the {transferred information} (\ref{non-unitary-qusi-Bell}) leads to a higher value, in comparison to the capacity given by Eq. (\ref{C-un-Bell}), i.e., $C^{Q,B}_{\Gamma}\textgreater C^{Q,B}_{\textmd {un}}$. In Figs. 4 and  5, we visualize the super dense coding capacity corresponding to unitary encoding and the  {transferred information} corresponding to the pre-processing $\Gamma$, Eqs. (\ref{C-un-Bell}) and (\ref{non-unitary-qusi-Bell}). 
In Fig. 4, the correlation degree is ${\mu}=0.2$, while we vary the noise parameter $p$. In Fig. 5, the noise parameter is $p=0.05$ and $\mu$ is varied
\newpage

\begin{center}
\begin{tabular}{c  }
\includegraphics[width=8 cm]{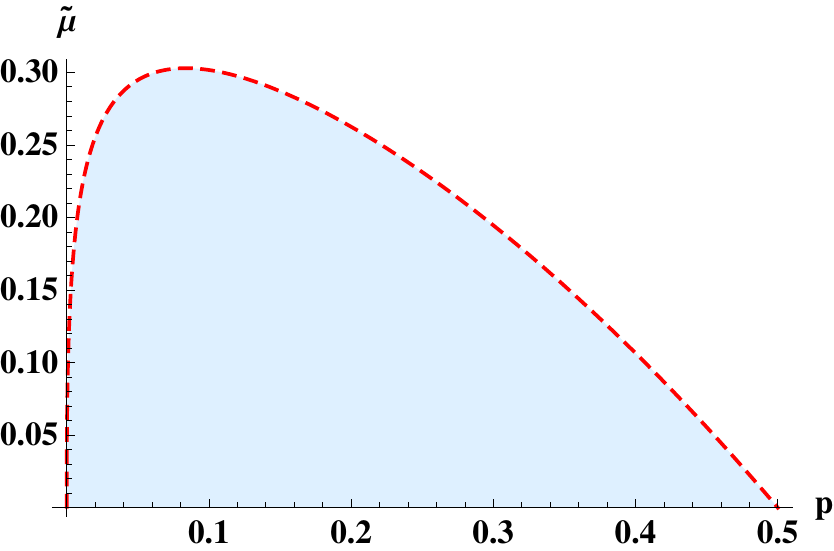}  \\
\end{tabular}
\label{mu-p}
\end{center}

Fig. 3. (Color online) The  dashed red  curve is the correlation degree $\tilde{\mu}(p)$ as a function of
the noise parameter $p$. The super dense coding capacity $C^{\textmd{Q,B}}_{\textmd {un}}$ and the transferred information $C^{\textmd{Q,B}}_{\Gamma}$ coincide for $\mu=\tilde{\mu}(p)$ (see main text). For $\mu \textless \tilde{\mu}(p) $, the
 blue (light gray)  area, the non-unitary pre-processing $\Gamma$ increases the super dense coding
capacity of a quasi-classical channel and a Bell state, in comparison to just
unitary encoding.

\vspace{2.5cm}

\begin{center}
\begin{tabular}{c  }
\includegraphics[width=8cm]{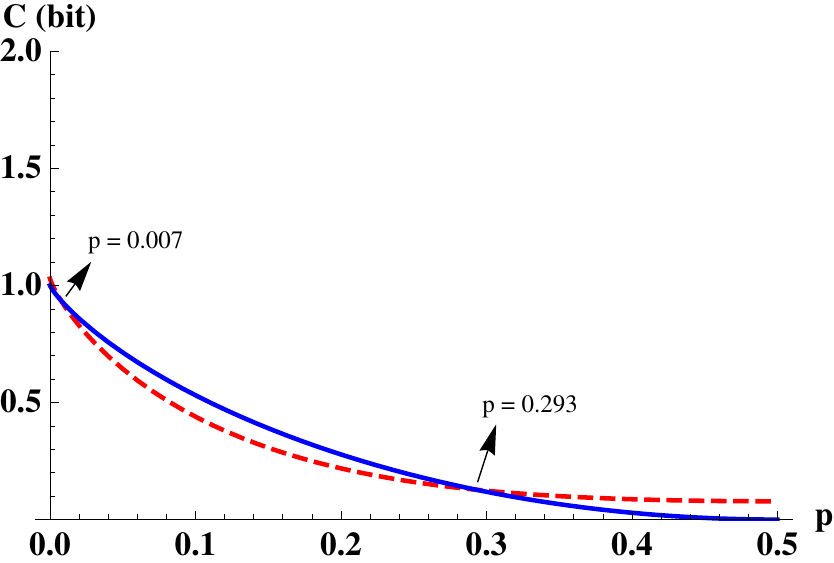}  \\
\end{tabular}
\label{mu-p}
\end{center}

Fig. 4. (Color online) Comparison between the super dense coding capacity (\ref{C-un-Bell}), and the transferred information (\ref{non-unitary-qusi-Bell}) as functions of the noise parameter $p$, with the correlation degree ${\mu}=0.2$. The  dashed red  curve corresponds to the capacity $C^{\textmd{Q,B}}_{\textmd {un}}$ given by eq. (\ref{C-un-Bell}), while the  solid blue curve represents  the transferred information $C^{\textmd{Q,B}}_{\Gamma}$ given by Eq. (\ref{non-unitary-qusi-Bell}). As we can see, for ${\mu}=0.2$, in the range of the noise parameter $ 0.007 \textless p \textless 0.293 $, we reach a higher capacity by applying the non-unitary pre-processing $\Gamma$, the  solid blue curve.


\begin{center}
\begin{tabular}{c  }
\includegraphics[width=7cm]{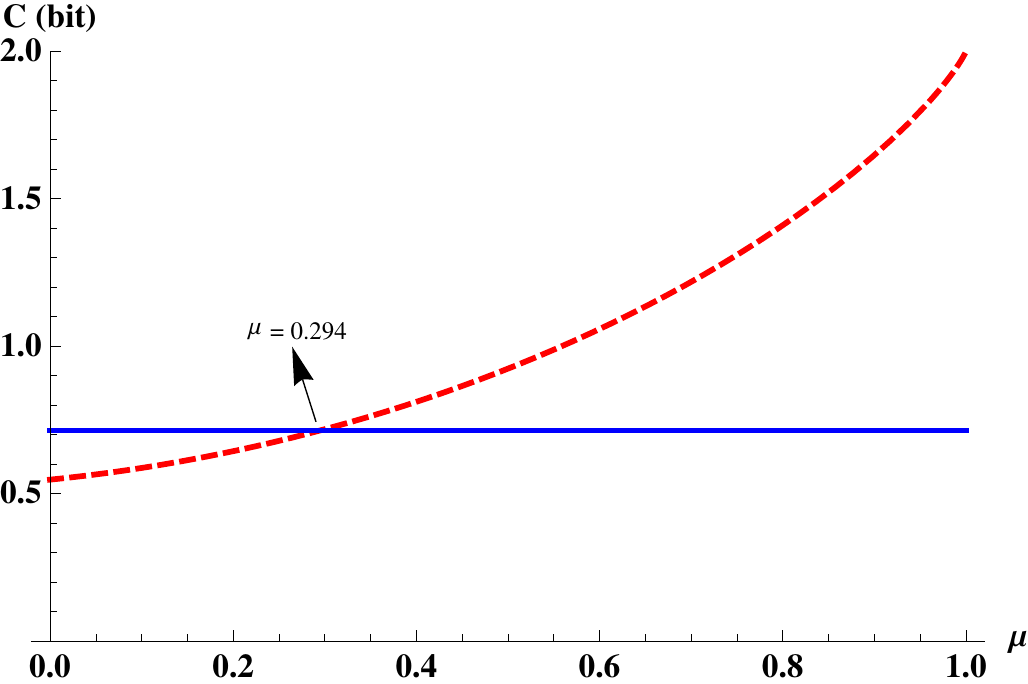}  \\
\end{tabular}
\label{mu-p}
\end{center}

Fig. 5. (Color online)  Another comparison between the super dense coding capacity (\ref{C-un-Bell}), and the transferred information (\ref{non-unitary-qusi-Bell}) as functions of the correlation degree $\mu$, with the noise parameter $p=0.05$. The  dashed red  curve corresponds to the capacity $C^{\textmd{Q,B}}_{\textmd {un}}$ given by eq. (\ref{C-un-Bell}). The  solid blue line represents the transferred information $C^{\textmd{Q,B}}_{\Gamma}$ given by eq. (\ref{non-unitary-qusi-Bell}). As we can see, for $p=0.05$, and  $ \mu \textless 0.294$, the non-unitary pre-processing $\Gamma$ is useful to enhance the capacity, compared to only unitary encoding.

\vspace{0.5cm}

\section{Conclusions \label{conc}}

In summary, we discussed the super dense coding protocol in the presence of a {correlated}  Pauli channel, 
considering both unitary and non-unitary encoding. Regarding unitary encoding, it was shown that the problem of finding the super dense coding capacity reduces to the easier problem of finding a unitary operator which is applied to the initial state such that it minimizes the von Neumann entropy after the
action of the channel. 
It was proven that for the two-dimensional quasi-classical channel and two-dimensional fully correlated Pauli channel with Bell states and Werner states as resources the unitary operator which minimizes the von Neumann entropy
is the identity. For those examples, the super dense coding capacities were 
analytically derived. We also showed that when considering non-unitary encoding, the optimal strategy is to apply a pre-processing before unitary encoding. If the map that minimizes the von Neumann entropy is known, we found  an expression for the super dense coding capacity. We also found an
explicit example of  non-unitary pre-processing $\Gamma$ which can improve 
the super dense coding capacity, in comparison to only unitary encoding, for 
a range of the correlation degree of the channel.  We also provided an example for which no better encoding than unitary encoding is possible. Therefore, based on the current results,  the usefulness of  non-unitary pre-processing in super dense coding depends on the quantum channel (e.g. its type, its noise parameter and correlation degree) and the resource state.


\begin{thebibliography}{1-10}
\bibitem{Bennett}  C. H. Bennett and S. J. Wiesner,  Phys. Rev. Lett. {\bf 69}, 2881 (1992).
\bibitem{hiroshima} T. Hiroshima, J. Phys. A  Math. Gen. {\bf 34}, 6907 (2001).
\bibitem{ourPRL} D. Bru{\ss}, G. M. D'Ariano, M. Lewenstein, C. Macchiavello, A. Sen(De), and  U. Sen, Phys. Rev. Lett. {\bf 93}, 210501 (2004).
\bibitem{Dagmar} D. Bru{\ss},  G. M. D'Ariano,  M. Lewenstein,  C. Macchiavello,  A. Sen(De), and  U. Sen, Int. J. Quant. Inform. {\bf 4}, 415 (2006).
\bibitem{ziman} M. Ziman, and V. Bu\v{z}ek, Phys. Rev. A {\bf 67}, 042321 (2003).
\bibitem{Bose-channel} S. Bose, Phys. Rev. Lett. {\bf 91}, 207901 (2003).
\bibitem{zahra-paper}  Z. Shadman, H. Kampermann, C. Macchiavello, and D. Bru\ss,  New J. Phys. {\bf 12}, 073042 (2010).
\bibitem{mp} C. Macchiavello and G. M. Palma, 
Phys. Rev. A {\bf 65}, 050301(R) (2002).
\bibitem{memory-quasi-chiara} C. Macchiavello, G. M. Palma, and S. Virmani, Phys. Rev. A {\bf 69}, 010303(R) (2004).
\bibitem{cerf} E. Karpov, D. Daems, and N. J. Cerf, Phys. Rev. A {\bf 74}, 032320
(2006).
\bibitem{Gordon} J. P. Gordon, in Proc. Int. School. Phys. "Enrico Fermi, Course XXXI", ed. P.A. Miles, 156 (1964).
\bibitem{Levitin} L. B. Levitin, Inf. Theory, Tashkent, pp. 111 (1969).
\bibitem{Holevo-chi-quantity} A. S. Holevo, \emph{ Information-Theoretical Aspects of Quantum Measurement}, Problems  Inform. Transmission, {\bf 9}:2, 110 (1973).
\bibitem{Holevo-capacity} A. S. Holevo, IEEE Trans. Inf. Theory {\bf 44}, 269-273 (1998).
\bibitem{Schumacher-Westmoreland} B. Schumacher, and  M. D. Westmoreland, Phys. Rev. A {\bf 56}, 131-138 (1997).
\bibitem{CPTP} M. Horodecki and M. Piani, quant-ph/0701134v2.
\bibitem{unitaryoptimal1} M. Horodecki, P. Horodecki, R. Horodecki,  D. W. Leung, and B. Terhal, Quantum Inf. Comput. {\bf 1}, 70 (2001).
\bibitem{unitaryoptimal2} A. Winter, J. Math. Phys. {\bf 43}, 4341 (2002).
\end {thebibliography}
\end{document}